\documentclass{aa}  
\usepackage{natbib,twoopt}
\usepackage{comment}
\usepackage[rightcaption]{sidecap}
\usepackage{hyperref} 
\hypersetup{
    colorlinks=true,
    linkcolor=red,
    filecolor=magenta,      
    urlcolor=cyan,
    pdftitle={Overleaf Example},
    pdfpagemode=FullScreen,
    citecolor=blue,
    breaklinks=true
    }
\bibpunct{(}{)}{;}{a}{}{,}             
\makeatletter
  \newcommandtwoopt{\citeads}[3][][]{\href{http://adsabs.harvard.edu/abs/#3}%
    {\def\hyper@linkstart##1##2{}%
     \let\hyper@linkend\@empty\citealp[#1][#2]{#3}}}
  \newcommandtwoopt{\citepads}[3][][]{\href{http://adsabs.harvard.edu/abs/#3}%
    {\def\hyper@linkstart##1##2{}%
     \let\hyper@linkend\@empty\citep[#1][#2]{#3}}}
  \newcommandtwoopt{\citetads}[3][][]{\href{http://adsabs.harvard.edu/abs/#3}%
    {\def\hyper@linkstart##1##2{}%
     \let\hyper@linkend\@empty\citet[#1][#2]{#3}}}
  \newcommandtwoopt{\citeyearads}[3][][]%
    {\href{http://adsabs.harvard.edu/abs/#3}
    {\def\hyper@linkstart##1##2{}%
     \let\hyper@linkend\@empty\citeyear[#1][#2]{#3}}}
\makeatother

\usepackage{adjustbox}
\usepackage{xcolor}
\usepackage{graphicx}
\usepackage{txfonts}
\usepackage{tabularx, blindtext}
\usepackage{rotating}
\usepackage{pdflscape}
\usepackage{inputenc}
\usepackage{caption}
\usepackage{longtable}
\usepackage{lscape}
\DeclareCaptionFormat{continued}{#1 (continued)#2#3\par}
\DeclareGraphicsExtensions{.jpg,.png, .pdf}
\usepackage[utf8]{inputenc}

\begin{document} 

\title{Over 200 globular clusters in the Milky Way and still none with super-Solar metallicity}
   \author{E. R. Garro \inst{1} 
          \and
          D. Minniti\inst{2,3,4}
               \and
             J.~G.~Fernández-Trincado\inst{5}
          }
   \institute{
   ESO - European Southern Observatory, Alonso de Cordova 3107, Vitacura, Santiago, Chile
   \and 
   Instituto de Astrofísica, Depto. de Ciencias Físicas, Facultad de Ciencias Exactas, Universidad Andrés Bello, Av. Fernandez Concha 700, Las Condes, Santiago, Chile
   \and
 Vatican Observatory, Vatican City State, V-00120, Italy
 \and
 Departamento de Fisica, Universidade Federal de Santa Catarina, Trinidade 88040-900, Florianopolis, Brazil
 \and
 Instituto de Astronomía, Universidad Católica del Norte, Av. Angamos 0610, Antofagasta, Chile
 }
  \date{Received: XX; Accepted: YY}
 
\abstract
   {
A large number of globular clusters in the Milky Way have been studied in recent years, especially in hidden regions such as those of the Galactic bulge.
    }
   {
The main goal of this work is to understand what we can learn if we include these new objects into the Milky Way globular cluster (GC) system that we know today. We compiled a catalog of 37 recently discovered globular clusters. Most of them are located in the Galactic bulge, but we also included some of the GCs for comparison. 
   }
   {
We used a range of distributions for investigating the Galactic GC system based on the metallicity, luminosity function, and age. We considered three samples. We first treated  the new GC sample separately from the known and well characterized GCs. Consequently, we merged these two samples, thereby upgrading the Milky Way GC system. Furthermore, we performed a comparison between our clusters sample and the field star population.   
   }
   {
We found a double-peaked distribution for the luminosity function, which shows an elongated faint end tail. Considering the "merged" sample, the luminosity function peaks at $M_{V}^{up} = -7.00 \pm 1.3$ mag and at $M_{V}^{up} = -4.1 \pm 0.48$ mag. The metallicity distributions also display a bimodality trend. In this case, we compare our new sample compilation with previously published ones, finding that the distributions are in good general agreement. We also constructed the metallicity distribution for the field star sample and, by comparing it with that  of the GCs, we learned that a high percentage of field stars show [Fe/H] $>0$; whereas we did not detect any GCs in the same metallicity range. To understand this inconsistency, we constructed the age-metallicity diagram for both samples, noting that the old and metal-poor population (age $\geq8$ Gyr and [Fe/H] $\leq -1.0$) is represented by Gcs, while the young and metal-rich population (age $<8$ Gyr and [Fe/H] $>-1.0$) corresponds to field stars. 
   }
  {
From the analysis of the GC luminosity function and metallicity distribution, we can conclude that many GCs, probably those that are very faint, have survived strong dynamical processes that are typical of the bulge regions. Moreover, we cannot exclude the possibility that some of them have been accreted during past merging events, especially the metal-poor component, whereas the metal-rich population may be related to the formation of the bulge and/or disk. Finally, the difference that we notice between the cluster and field star samples should be explored in the context of the evolutionary differences among these two stellar populations.
   }
\keywords{Galaxy: bulge – (Galaxy:) globular clusters: general - Galaxy: formation - Galaxy: evolution}
  
   \maketitle
    
\section{Introduction}
\label{Introduction}
In a cosmological context, the hierarchical galaxy formation process predicts that galaxies and their halos have a large number of progenitors at the formation redshift ($z\sim 3$) of globular clusters (GCs, \citealt{DeLucia2007,Lamers2017}). With the advent of new observations from {\it James Webb} Space Telescope, new theories may be formulated and new observational discoveries can be established, which may change, confirm, or improve our knowledge. For instance, recent works have uncovered (mini-)quenched galaxies at redshifts of  $z=4.4$ \citep{2023arXiv230602470L} and $z=7.3$ \citep{2023arXiv230214155L}; thus, we cannot exclude the fact that GCs have been formed as well at these redshifts. Also, \cite{Mowla_2022} found compact red companions to the Sparkler galaxy, which may be the first evolved GCs seen at $z_{spec}\approx 1.4$. They infer formation redshift of $z_{form}\sim 7-11$ for these candidates, corresponding to ages of $\sim 3.9 - 4.1$ Gyr at the epoch of observation and a formation time just $\sim 0.5$ Gyr after the Big Bang. Therefore, as these authors reported, if these ages are confirmed, at least some GCs appear to have formed contemporaneously with the large-scale reionization of the intergalactic medium, hinting at a deep connection between GC formation and the initial phases of galaxy assembly. Nevertheless, there is a long way to go before we can at least confirm the existence of those primordial GCs and begin constructing new models that describe these compact objects. This is especially true because their exact formation redshifts are somewhat uncertain for these unresolved and red objects, with no spectroscopic data yet available to offer clues to the real phase of these objects.

Both simulations (e.g., \citealt{Tonini2013, Kruijssen2015}) and observations (e.g., \citealt{Horta2022}) have shown that much of the metal-poor GC population in massive galaxies was mostly accreted from cannibalized dwarf galaxies, whereas the metal-rich population formed preferentially in situ\footnote{We use "in situ” to refer to the GCs that formed during the initial collapse of the central galaxy, and "ex-situ" for those one that are accreted during the subsequently mergers.} or also likely during merging events \citep{Hughes2019,Garro2021b, Hasselquist2021}. In this context, debris from accreted GCs and dwarf galaxies is expected to populate halo regions as well as the central regions of the MW \citep{Das2020, Fern_ndez_Trincado_2019, Fern_ndez_Trincado_2021}. However, the number of confirmed merging footprints in the inner regions is still smaller than those found in the Galactic halo. For that reason, we want to investigate the Galactic bulge to collect clues to its formation and evolution history.\\
  
Actually, there are many queries that remain unanswered regarding the bulge. For instance, it is not completely clear if the bulge is a simple classical bulge, formed from a monolithic collapse, and then grown up hierarchically by merging of galaxies \citep{Guedes_2013}, or it was formed by secular evolution  \citep{2004ARA&A..42..603K}, so it is part of the peanut structure corresponding to the old Galactic thick disk. This would imply that the MW is a pure thin+thick disk galaxy, with only a possible limited contribution by a classical bulge  \citep{DiMatteo2015}. Such a mix would represent a compound bulge \citep{10.1111/j.1365-2966.2005.08872.x,10.1093/mnras/stu2376, 2014A&A...562A..66Z}. Secondly, there are also queries about its evolution, such as how many accretions occurred, how the star formation history and so the chemical evolution proceeded, and so on.\\

The main aim of this paper is to take advantage from bulge GC populations, which can be used as archaeological probes to reconstruct the formation and evolution of the MW. However, such regions as the Galactic bulge are very challenging to study, because of the high stellar crowding and differential reddening, which make the identification of new GCs a complicated task. Wide-field near-infrared (NIR) photometry such as the VISTA Variables in the Vía Láctea (VVV; \citealt{2010NewA...15..433M,2012A&A...537A.107S}) survey and its extension (VVVX; \citealt{2018ASSP...51...63M}) offers a huge advantage to detect several sparse and faint GCs heavily obscured by interstellar dust in the bulge region. Indeed, numerous ($\sim 300$) GC candidates have recently been discovered in these zones \citep{Minniti_2017,Minniti2021_M48, Minniti2021_cl160,2022A&A...657A..67D, Garro2021a, Garro2022_Garro02}. 
It seems that the number of known MW GCs (in total $N_{GC} \sim 170$ - \citealt{Vasiliev2021b}), but in particular the MW bulge GCs ($N_{GC} \sim 40$ - \citealt{Barbuy2018}, or maybe more as suggested by \citealt{2016PASA...33...28B}) is too small if we take into account the mass of the MW and if we compare that number with similar systems, such as M31. In this context, for example, \cite{Abdullah2018SpecificFO} (see their Figure 6) 
showed that the MW and M31 specific frequencies should be similar in their range of masses ($M/M_{\odot} = 10^{10} - 10^{11}$). Therefore, this may imply that the census of the MW clusters is still incomplete \citep{Minniti_2017}. Indeed, 
searching and analysing new GCs in the MW bulge is tricky. They appear as round concentrations of stars, showing up as overdensities above the background in the optical and NIR images, but sometimes they are not real clusters but asterisms, background fluctuations, or possible streams of previously disrupted GCs. At the distance of the Galactic bulge, their typical half-light radii would range between 2 pc ($0'.80$) for the smallest ones and 10 pc ($4'.20$) for the largest ones.  Additionally, since many of them are very faint and small objects, they are experiencing destruction processes, due to the fact that the deep potential well of the Galactic bulge strengthens the efficiency of such dynamic processes \citep{2014A&A...566A...6H,2018A&A...620A.154K}. Hence, this is why the Galactic bulge is referred to as the elephant graveyard \citep{Minniti_2017}.\\

Hence, we want to employ both known and well characterized GCs and the recently discovered ones for investigating the inner regions, using different tools: the metallicity distribution function (MD), the luminosity function (LF), age distributions, and the kinematics information. Additionally, we prefer expanding this analysis to the bulge field star (FS) populations, since it is expected that a small fraction of bulge FSs (proposed to be $\sim 5-10\%$ -- \citealt{Kruijssen2012,Adamo2015}) is born in bound systems, and then it is lost due to both merging events and/or destruction/dissolution processes (stellar evolution, two-body relaxation, tidal evaporation, and tidal shocks -- \citealt{Kruijssen2011,Madrid2017}). Consequently, a comparison between these two populations may help us to understand and constrain the formation and evolution of the inner Galaxy. Nevertheless, these two populations are usually treated separately. \\

With respect to what has been done thus far, different studies have reported that the Galactic bulge shows multiple peaks in its MD,  features that are usually associated with different stellar populations \citep{Ness_2013,Bensby2017,Rojas_Arriagada2020}, coexisting in the innermost regions. The existence of a metallicity gradient is demonstrated adopting different tracers \citep{1995MNRAS.277.1293M}.  \citet{Gonzalez2013} using the VVV photometry confirmed that metal-rich stars ([Fe/H] $\sim 0$) dominated the inner bulge in regions closer to the Galactic plane ($|b| < 5$), while at larger scale heights, the mean metallicity of the bulge red giant population becomes significantly more metal-poor. \citet{Ness_2013} and \citet{10.1093/mnras/stx947} found that the metal-rich population is dominant at low latitudes, but drops off rapidly with increasingly latitude. The metal-poor population instead follows the opposite trend, increasing toward high Galactic latitudes. The intermediate-metallicity population dominates everywhere except near the disk, where the metal-rich population dominates. On the other hand, works by \citet{Barbuy2018a} and \citet{Savino2020} found a predominance of moderately metal-poor populations in the innermost Galactocentric regions, which extends to large distances above the Galactic plane.  This is probably related to the fact that the old and intermediate age RR Lyrae populations were taken into account and they are not expected to exhibit a strong peanut shape, as red clump stars \citep{2015A&A...583L...5G}, since they should be older than the formation of the bar itself \citep{10.1093/mnras/stx947}. On the other side, if we take into account GCs, we find the metal-rich ones especially in the inner regions of the Galaxy, while the metal-poor one in the halo. \\
We may note that the meanings of "metal-poor" and "metal-rich" are not strictly defined and depend on the different contexts where these terms are used. For instance, for the "metal-rich" FSs we often refer to solar metallicity ([Fe/H] $\approx 0$), while for the "metal-rich" GCs, we consider values of [Fe/H] $> -0.5$, which might be referred to as "intermediate-metallicity" for the field overall. Therefore, in the present work, we broadly define "metal-poor" as [Fe/H] $<-1.0$, whereas "metal-rich" if [Fe/H]~$\geq -1.0$, both for FSs and GCs. Moreover, we refer to "super metal-rich" or "super-Solar" to indicate those populations with [Fe/H] $\geq 0$.
Various nomenclatures are often used in the literature, such as "moderately metal-poor" (which may correspond to $-2.0 \lesssim$ [Fe/H] $\lesssim -1.0$) or "intermediate metallicity" (which may be equivalent to $-1.5\lesssim$ [Fe/H] $\lesssim -0.6$); however, this depends once again on the populations involved. We prefer not to use this nomenclature.\\

Today, we know of $\sim 170$ well-characterized GCs in the MW \citep{Vasiliev2021b}.  Broadly, the Galactic GC system shows a bimodal metallicity distribution, which indicates that in the MW coexist two major subpopulations, but always the metallicity is typically sub-Solar ([Fe/H]~$<0$). In spite of that, there are some GCs that may show [Fe/H] $>0$. This is not completely demonstrated, and the misgivings are strongly related to the abundance scale's calibrations adopted. Therefore, we cannot ignore them. For example, \cite{Carretta2009} suggested a [Fe/H]~$>+0.40\pm 0.09$ for Liller~1 GC. Recently, \cite{Dalessandro_2022} and \cite{2023ApJ...951...17C} showed a bimodal iron distribution, with a sub-solar component ([Fe/H] $=-0.48$ dex with $\sigma=0.22$, and age $\approx 12$ Gyr) and a super-solar component ([Fe/H] $=+0.26$ dex with $\sigma=0.17$, and age $\approx 1-3$ Gyr) for Liller~1. For this purpose, they suggested that Liller 1 experienced three main bursts and a low, but constant, activity of star formation over the entire lifetime. Also, NGC~6528 may have a super-solar metallicity as found by \cite{Carretta2009} ([Fe/H] $=+0.07\pm 0.08$) and \cite{2017A&A...601A..31L} ([Fe/H] $=+0.04\pm 0.07$), but in both cases, the uncertainties are greater than the value itself; therefore, we cannot definitively conclude either way whether we indeed have a case of a super metal-rich GC. It is clear that the analysis of the bulge MD is very complex, since this strongly depends on target selection, tracers, and analysis methods. \\

Moreover, the $\alpha$-elements are efficient tracers of Galactic evolution, especially when [$\alpha$/Fe]-[Fe/H] diagrams are used. An $\alpha$-enhanced plateau at low [Fe/H] is typical both for the MW and external galaxies, this is due  to the fast enrichment from massive ($8 <M/M_{\odot}<40$) stars that explode as type II supernovae (SNII), while a following decline is expected once low-mass (in binary systems with $1.4 <M/M_{\odot}<8$) stars begin to explode as type Ia supernovae (SNIa). This creates the "knee" point, which can move to higher metallicity for higher star formation rates \citep{Matteucci1986}. For example, \citet{Queiroz2021} found an apparent chemical bimodality in key abundance ratios, such as [$\alpha$/Fe], [C/N], and [Mn/O], which probe different enrichment timescales, indicating a star formation gap between the $\alpha$-rich and $\alpha$-poor populations. 
They also found signatures of the bar, a thin inner disk, a thick inner  disk, and a broad metallicity population with large velocity dispersion indicative of a pressure-supported component. Furthermore,  they found a group of counter-rotating stars that could be an accreted metal-poor population originated during a gas-rich accretion phase in the early formation of the bulge, or of a clumpy star formation during the earliest phases of the early disk that migrated into the bulge \citep{Fiteni2021}. In summary, many observations suggest that the formation of the boxy bulge occurred later, originating from the vertical instability of the Galactic bulge, formed via secular evolution from the stellar disk.  In this scenario, the old and metal-poor stellar population dominates at higher scale heights due to the non-mixed orbits of originally hotter thick disk stars.  \citet{Lian2020} derived three-phase star formation history of the MW bulge, which consists of an initial starburst, followed by a rapid star formation quenching episode and a lengthy, quiescent secular evolution phase. In fact, they found that the metal-poor, high-$\alpha$ bulge stars ([Fe/H] $< 0.0$ and [Mg/Fe] $>0.15$) were formed rapidly ($< 2$ Gyr) during the early starburst. The density gap between the high- and low-$\alpha$ sequences is due to the quenching process. The metal-rich, low-$\alpha$ population ([Fe/H] $>0.0$ and [Mg/Fe] $<0.15$) then accumulates gradually through inefficient star formation during the secular phase.  This is qualitatively consistent with the early star formation history (SFH) of the inner disk.  Additionally, this explains why there is a fraction of young stars with age $ < 5$ Gyr in the bulge.  Indeed, we know that the bulge mean age is about 10 Gyr,  with a lower-limit of $\sim 5$ Gyr \citep{2011ApJ...735...37C}, where well-mixed clusters and stellar populations coexist.  The typical age of GCs in the innermost regions is older than 10 Gyr \citep{1996AJ....112.1487H,Minniti1995}, however GCs with intermediate age ($\approx 7-9$ Gyr) are found by \citet{2022A&A...658A.120G} and \citet{2022A&A...659A.155G}, indicating that these objects are either real members of the bulge or intruders originating from the halo \citep{Forbes_Bridges2010,2019A&A...630L...4M,RomeroColmenares2021}. Moreover,  spectroscopy of microlensed stars by \citet{Bensby2017} found a very wide age distribution for the bulge, with several star forming episodes at 3, 6, 8, and 11 Gyr. Also, they linked the ages to the metallicity of the microlensed stars.  At [Fe/H] $\approx -0.5$ all stars appear to be older than 10 Gyr.  At higher metallicities, the age ranges spans from the youngest one with $\sim 1$ Gyr to the oldest with $\sim 12-13$ Gyr.  Additionally, the authors indicated that their age becomes younger with increasing metallicity.  Below [Fe/H] $\approx -0.5$ essentially all stars are old, below solar but more metal-rich than $-0.5$ dex, the fraction is around 20\%, and above solar metallicity more than one third of the stars are younger than 8 Gyr. Moreover, \citet{Bensby2017} found that Galactic bulge properties (MFD, star formation episodes, abundances trends) are similar to those of the MW thin and thick disks. However, it seems that the bulge star formation rate was slightly faster than that of the local disk, as also proposed by \citet{Lucertini2022}.  Finally, their results again strengthen the observational evidence that supports the idea of a secular origin for the Galactic bulge, formed out of the other main stellar populations present in the central regions of our Galaxy. \\

To summarize, it  really is difficult to understand which of the three theories (hierarchical scenario, secular evolution, or a combination of these two) best explains the formation and evolution of the bulge, as well as that of the entire galaxy. Therefore, we prefer to use the GC population in our attempts to provide constrains, since these objects trace the Galaxy’s properties from the innermost regions toward the outskirts. In particular, we analyzed numerous low-luminosity GCs, which probably survived strong dynamical processes. We employ them in this paper to see what we can learn if we include this sample in the typical MW GC system.

\section{Sample of selected bulge GCs}
\label{targetselection}
\begin{figure*}[htpb]
\includegraphics[width=9cm, height=7cm]{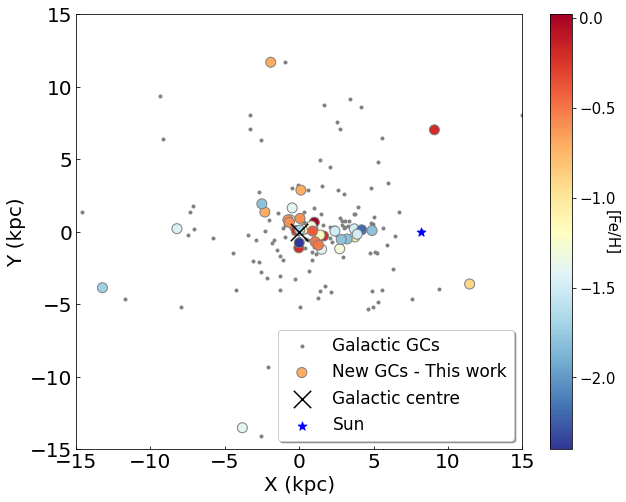} 
\includegraphics[width=9cm, height=7cm]{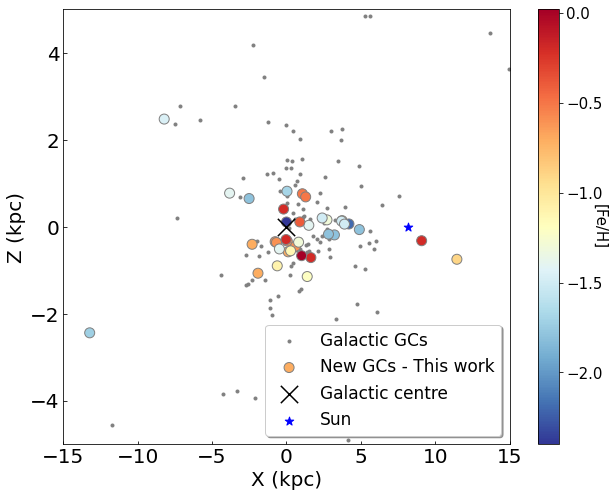} 
\caption{Galactocentric distribution (XY on the left, and XZ on the right) of the GCs analyzed in the present work (colored points), highlighting clusters in the inner regions of the bulge. The known Galactic GCs positions are drawn in grey points. X measured the system's "east-west" location, Y measured its "north-south" location, and Z measured its distance above or below the galactic plane. The marker's color changes depending on the cluster's metallicity, as indicated by the color bar. The black cross represents the position of the Galactic centre, while the blue star indicates the position of the Sun.}
\label{fig:Positions}
\end{figure*}

As mentioned previously, many new GCs are recently discovered and photometrically analyzed both in NIR and optical wavelengths in the MW bulge and disk. We have cataloged these objects in Table \ref{table:Table1}, where we listed their main physical parameters: positions, mean cluster proper motions (PM), heliocentric and Galactocentric distances, metallicities, ages, extinction, and integrated luminosities. We also list the references from which we have taken all the information. The majority of the clusters are located inside the Galactic bulge at $R_{G} \lesssim 3$ kpc; other clusters, namely Garro~01,  Riddle 15, and Gaia 2, are situated in the disk at $R_{G}>11 $ kpc. Notably, clusters, located in the innermost regions, are very obscured and heavy embedded by the dust, where the differential reddening and extinction ($0.1 <A_{Ks}< 1.0$ mag, equivalent to $0.9 < A_{V} < 9.1$ mag - see Table \ref{table:Table1}) can play an important role both for the detection and the characterization of these new objects. Moreover, Figure \ref{fig:Positions} displays the position of the GCs analyzed in Galactocentric coordinates, and we highlight the metallicity (changing the color of the points) in order to show that both metal-poor and metal-rich GC populations are present inside the bulge (see also Section \ref{MDF}). The spatial distribution of the new sample seems to be asymmetric, preferentially toward $X>0$ kpc. This is due to the fact that most clusters present in our compilation were discovered in the VVV-VVVX footprints, but this does not mean that on the opposite side  ($X<0$ kpc), other small GC candidates could live. Likely, they have not been detected yet. \\

\begin{figure}[htpb]
\centering
\includegraphics[width=8cm, height=8cm]{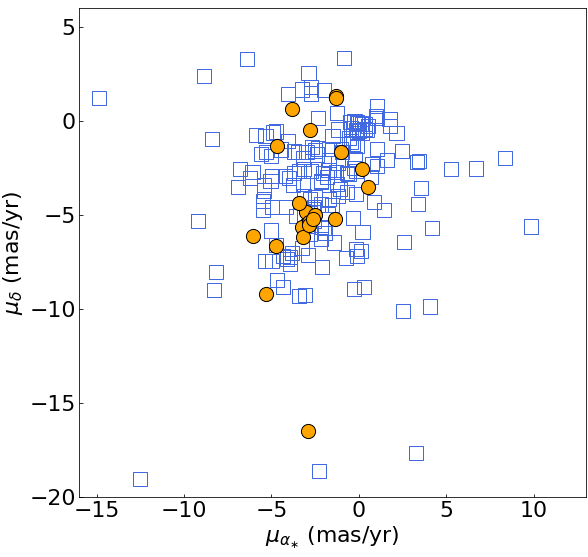} 
\caption{Vector PMs diagram for our GC sample. It demonstrates that the kinematics of the GCs here studied (orange points) are consistent with the Galactic GCs from \citet{Vasiliev2021b} (blue open squares).}
\label{fig:Kinematics}
\end{figure}

Furthermore, we selected the clusters depending on their mean cluster PMs (where available), and we compare our sample with well-known Galactic GCs studied by \citet{Vasiliev2021b}.  As shown in Figure \ref{fig:Kinematics}, our PM values are indeed consistent with the Galactic GCs sample. We can easily distinguish VVV-CL160 with $(\mu_{\alpha_{\ast}}, \mu_{\delta} = -2.90, -16.47$) mas yr$^{-1}$, which is separated from the bulk of the points. The origin of VVV-CL160 is still debated \citep{Minniti2021_cl160}, since it was found that its orbit is one of the most eccentric between the Galactic GCs \citep{Garro2023}. \\

It is important to mention some important limitations of our work. First, most of the clusters in this new catalog have been only identified photometrically as GC candidates, therefore they deserve a detailed spectroscopic analysis for deriving robust [Fe/H] estimates as well as radial velocities and chemical abundances ([X/H]), in order to further confirm their real nature.  For example,  \citet{Lim2022} carried out a spectroscopic analysis of 15 and 10 target stars near Camargo 1103 and Camargo 1106, respectively, two star clusters that have been recently reported as metal-poor GC candidates in the bulge \citep{Camargo_2018}. They did not find any clear evidence of a grouping in radial velocity-metallicity space that would indicate the characterization of either object as metal-poor GCs. Yet, they cannot completely rule out a low probability that they only observed non-member stars. Another example may be represented by Garro 01, since its classification is ambiguous. It was classified as GC with [Fe/H] $= -0.7$ and age $\approx 11$ Gyr (as written in Table \ref{table:Table1}) by \cite{Garro_2020} and recently as possible open cluster with [Fe/H]~$=-0.3$ and age $=4$ Gyr by \cite{2023arXiv230406904P}. However, they cannot confirm the Garro 01 cluster nature either. Here, we assume that all these targets (Table \ref{table:Table1}) are genuine GCs. Second, incomplete color-magnitude diagrams (CMDs) that do not reach the main sequence turnoff point (MSTO),  affects the age and luminosity estimates. In particular, in many cases ages are derived using the vertical method (see for example the work by \citealt{2022A&A...658A.120G}) or identifying  RR Lyrae stars belonging to the clusters, thus fixing a lower limit, but indicating large error bars (Table \ref{table:Table1}). This also implies that we cannot aptly classify them as (young) GCs or old open clusters (OCs), even if a net separation is clearly established between these two classes. 

A third caveat is represented by the fact that many of these clusters (considered in the present work) are located in very obscured regions, where the differential reddening, high extinction and stellar crowding may complicate the analysis, introducing uncertainties in the derivation of the main physical parameters. However, these problems are highly discussed by the authors of the papers (see Table \ref{table:Table1},  last column). Finally, taking into account the errors in the photometry, we can consider that this first work on the newly updated MW GC system can offer further clues about the GC system itself, as well as indications about the MW's history. \\

This study includes the information collected to date, and the sample considered here contains both the known and well-characterized GCs and those from recently studied and confirmed as high-probability GCs (Table \ref{table:Table1}). In this way, we are upgrading the MW GC system, therefore we expect that this has an impact on the MW GC luminosity function (LF, see Sect. \ref{MWGCLF}) and the metallicity distribution function (MD, see Sect. \ref{MDF}).

\begin{table*}
\centering 
\renewcommand{\arraystretch}{1.4}
\caption{Catalog of the 37 recently discovered GCs in the MW.  We list (where available) their main physical parameters: position,  mean cluster PMs, extinction in $K_s$ band, distances, metallicity, age, total luminosity, and their references.}
\begin{adjustbox}{max width=\textwidth}
\begin{tabular}{lcccccccccccccccccccl}
\hline\hline
Cluster ID      &  l         &  b        & $\mu_{\alpha_{\ast}}$   &  $e_{\mu_{\alpha_{\ast}}}$  &   $\mu_{\delta}$  &  $e_{\mu_{\delta}}$  & $A_{Ks}$ & $e_{A_{Ks}}$  & $D_{\odot}$  & $e_{D_{\odot}}$  & $R_G$ &  [Fe/H]    & $e_{[Fe/H]} $  & Age    & $e_{Age}$   & $M_{Ks}   $ & $e_{MKs}$   & $M_V $  & $e_{M_V}$  & Reference \\
  & [deg] & [deg] & [mas yr$^{-1}$] & [mas yr$^{-1}$]&[mas yr$^{-1}$] &[mas yr$^{-1}$] & & & [kpc] & [kpc] & [kpc] &[dex] & [dex] &[Gyr] &[Gyr] &[mag] &[mag] &[mag] &[mag] &\\ 
\hline
Garro 01     &  310.828 &  $-3.944$   &  $-$4.68  &   0.47    &  $-$1.35   &   0.45 & 0.15 & 0.01 &  15.5  & 1.0  & 11.2   &  $-$0.7   & 0.2   & 11.0 & 1.0  & $-$7.76    &  0.5   & $-$5.26  & 1.0  & \tablefootmark{(a)} \tablefootmark{(z)} \\
Garro  02  &  12.048 & 1.646 & $-$6.07 & 0.62 & $-$6.15 & 0.75 &0.79 & 0.04 & 5.6 & 0.8 & 2.90 & $-$1.3 & 0.2 & 12.0 & 2.0 & $-$7.52 & 1.2 & $-$5.44 & 0.8 &         \tablefootmark{(b)}\\
Patchick 99    &  2.488   &  $-6.145$   &  $-$2.98  &   1.74    &  $-$5.49   &   2.02 & 0.09 & 0.01   &  6.6   & 0.6  & 2.81   &  $-$0.2   & 0.2   & 10.0 & 2.0  & $-$7.00    &  0.6   & $-$5.20  & 1.0  &  \tablefootmark{(c)}\\
FSR 0009 & 1.856 & $-$9.529 & $-$1.39 &1.10 &$-$5.22 & 0.99 & 0.11 & 0.03& 6.9 & 0.2 & 1.82& $-$1.2 & 0.3 & 11.0 & 2.0 & $-$5.80 & 0.7 &$-$3.40 &1.0 & \tablefootmark{(d)}\\ 
FSR 1775 & 354.546 & $-$5.779 & $-$3.00 & 0.80 & $-$5.53 & 0.73 & 0.16 & 0.01 & 8.9 & 0.2 & 1.37 & $-$1.1 & 0.2 & 10.0 & 2.0 & $-$8.00 &1.7 &$-$5.60&1.5 & \tablefootmark{(d)}\\
FSR 1767 & 352.601 & $-$2.166 & $-$3.02 & 0.50 & $-$4.85 &0.50 &0.28 &0.03 & 10.6 & 0.2 & 2.70 & $-$0.7 & 0.2 & 11.0 &2.0 & $-$8.40 & 1.5 & $-$6.30 &1.3 &\tablefootmark{(d)}\\
VVV-CL131 & 354.722 & $-$2.170 & $-$3.24 & 0.81 & $-$5.65 & 0.07 & 0.23 & 0.03 & 9.0 &0.5 & 1.17& $-$0.6 & 0.2 & 10.0 & 3.0 & $-$8.20 & 1.5 & $-$5.90 & 1.3&\tablefootmark{(d)}\\
VVV-CL143 & 355.788 & $-$2.319 & $-$3.18 &0.91 & $-$6.17 & 0.85 &0.21 &0.04  & 8.9 & 0.5 & 1.00 & $-$0.6 & 0.2& 10.0 & 3.0 & $-$8.20 & 1.3 & $-$5.90 & 1.2 &  \tablefootmark{(d)}\\
VVV-CL160  & 10.151 & 0.302 & $-$2.90 & 1.28 & $-$16.47 & 1.31 & 0.73& 0.03 & 6.8 & 0.5 & 1.92 & $-$1.4 & 0.3 & 13.0 & 2.0 & $-$7.90 & 1.5 & $-$5.50 &1.2&   \tablefootmark{(d)}  \tablefootmark{(e)}\\
ESO 393-12 & 353.514 & $-$2.284 & $-$2.86 & 9.47 & $-$5.39 & 0.44 &0.23 & 0.02 & 8.2 & 0.4 & 0.98 & $-$0.6 & 0.2 & 10.0 &2.0 & $-$7.70 & 1.5 & $-$5.30 &1.3 &\tablefootmark{(d)} \\
ESO 456-09 & 357.882 & $-$3.339 & $-$3.41 & 0.71 & $-$4.36 & 0.75 & 0.18&0.03  & 7.6 & 0.4 & 0.81 & $-$0.6 & 0.2  &10.0 & 2.0 & $-$8.30 & 1.5 & $-$6.00 & 1.3 &\tablefootmark{(d)} \\
Kronberger 49 & 7.627 & $-$2.012 & $-$2.84 & 0.69 & $-$5.52 & 0.71 & 0.30 & 0.03 & 8.3 & 0.5 & 1.14 & $-$0.2 & 0.2 & 11.0 & 2.0 & $-$8.50 & 1.5 & $-$6.70 & 1.3 & \tablefootmark{(d)}\\
Patchick 125   &  349.756 & 3.426 & $-$3.85 & 0.50 & $+$0.64 & 0.39&0.33 & 0.01 & 10.9 & 0.5 & 3.23 & $-$1.8 & 0.2 & 14.0 & 2.0 &$-$6.10 & 0.8 & $-$3.80 & 1.0 & \tablefootmark{(f)}\tablefootmark{(g)} \tablefootmark{(h)} \tablefootmark{(z)}\\
Patchick 126 & 340.381 & $-$3.826 & $-$4.75 & 0.46 & $-$6.68 & 0.62& 0.44 & 0.01 & 8.6 & 0.4 & 2.94 & $-$0.7 & 0.3 & 8.0 & 4.0 & $-$5.56 &0.8 & $-$3.50 & 1.0 & \tablefootmark{(f)}\\  
Riddle 15      &  48.355 & 2.455 & $-$1.03 & 0.32 & $-$1.64 & 0.27& 0.52 & 0.01  &18.10 & 0.5 & 14.06& $-$1.4 & 0.2 & 13.0 & 3.0 & $-$7.60 & 0.8 & $-$6.20 &1.0& \tablefootmark{(f)}\\
Ferrero 54      & 262.803 & $-$2.571  & $-$1.33 & 0.27 & $+$1.31 & 0.34& 0.60&0.01  & 7.10  & 0.4 & 11.5 & $-$0.2 & 0.2 & 11.0  & 2.0 &$-$5.87 & 1.7 & $-$4.10 & 1.5 &\tablefootmark{(f)}\\
Gaia 2  & 132.155 & $-$8.730  &$-$1.31 & 0.18 & $+$1.21  & 0.19& 0.10 & 0.002 & 4.91 & 0.5 & 12.03  & $-$0.9 & 0.2 & 10.0 & 1.0 & $-$5.40 & 2.0 & $-$3.9 & 1.7 &\tablefootmark{(f)}\\
VVV-CL001      &  5.267   &  $ 0.780$   &  $-$3.41  &   0.50    &  $-$1.97   &   0.5& 0.65 & 0.03   &  8.23  & 0.4  & 0.76   &  $-$2.4   & 0.24  & 11.9 & 3.0  &            &        &          &      &\tablefootmark{(i)} \tablefootmark{(l)} \tablefootmark{(m)}\\
FSR 1776       &  354.720 &  $-5.249$   &  $-$2.3   &   1.10    &  $-$2.6    &   0.8 & 0.12&   &  7.24  & 0.5  & 1.39   &  $+$0.02  & 0.14  & 10.0 & 2.0  &            &        &          &      & \tablefootmark{In)}\\
Gran 1         &  -1.233  &  $-3.978$   &  $-$8.10  &           &  $-$8.01   &         & 0.24 & & 7.94    &      & 0.64   &  $-$1.13  & 0.06      & 10.0 & 2.0  &            &        & $-$5.46  &      &  \tablefootmark{(h,y)}\\
Gran 2         &  -0.771  &  $8.587 $   &  $+$0.19  &           &  $-$2.57   &      &0.26 &    &  16.6  &      & 8.58   &  $-$1.46  & 0.13  &      &      &            &        & $-$5.92  &      &  \tablefootmark{(h,y)}\\
Gran 4             &  10.198  &  $-6.388$   &  $+$0.51  &           &  $-$3.51  & & 1.40 &  & 21.9  & &   13.7   &$-$1.72& 0.32  &10.0  &  &   &     &   $-$6.45 & &   \tablefootmark{(h,y,z)}\\
Gran 5         &  4.459   &  $1.838 $   &  $-$5.32  &           &  $-$9.20   &  &0.43 &       &  4.47  &      & 3.76   &  $-$1.56  & 0.17  &      &      &            &        & $-$5.95  &      &  \tablefootmark{(h,y)}\\
FSR 19         &  5.499   &  $6.071 $   &  $-$2.50  &   0.76    &  $-$5.02   &   0.47  & 0.19 & 0.07& 7.2   & 0.7  & 1.48   &  $-$0.5   &       & 11.0 &      & $-$7.72    &        & $-$4.62  &      & \tablefootmark{(o)}\\              
FSR 25         &  7.534   &  $5.649 $   &  $-$2.61  &   1.27    &  $-$5.23   &   0.74  & 0.27 & 0.01& 7.0   & 0.9  & 1.73   &  $-$0.5   &       & 11.0 &      & $-$7.31    &        & $-$4.21  &      & \tablefootmark{(o)}\\ 
FSR 1758       &  349.217 &  $-3.292$   &  $-$2.85  &   0.05    &  $+$2.47   &   0.05  & 0.26 & 0.07& 8.8  & 2.1  & 1.79   &  $-$1.4   & 0.05  & 11.6 & 1.3  &            &        &          &      & \tablefootmark{(p)}\\
Minni 48       &  359.35  &  $ 2.79 $   &  $-$3.50  &    0.5    &  $-$6.0    &    0.5  &0.45 & 0.05&  8.4  &  1. 0 &        &  $-$0.2   &  0.30 & 10.0 & 2.0  & $-$9.04    &   0.7  &          &      & \tablefootmark{(q)}\\
FSR 1716 (Minni 22)       &  356.828 &  $-2.729$   &           &           &            &         &  0.38& 0.02& 7.4  &  0. 3 &        &  $-$1.3   &  0.3  & 11.2 & 2.0  &            &        & $-$6.2   &  0.5 &  \tablefootmark{(r)} \\
VVV-CL002      &  359.559 &  $ 0.889$   &  $-$2.80  &           &  $-$0.5    &         & 0.98 & 0.29& 7.3  &  0. 9 &  0.70   &  $-$0.54   &  0.2  &  >6.5 &      &            &        & $-$3.4   &      &  \tablefootmark{(s)} \tablefootmark{(t)} \tablefootmark{(w)} \\
Camargo 1102   &  359.145 &  $ 5.734$   &           &           &            &         & 0.40 & 0.11 & 8.2  &  1. 2 &  0.85  &  $-$1.7   & 0.2   & 13.3 & 1.0  &            &        & $-$6.3   &  0.6 & \tablefootmark{(u)}\\
Camargo 1103   &  5.604   &  $-2.121$   &           &           &            &         & 0.73 & 0.11& 5.0  &  0. 8 &  3.03  &  $-$1.8   & 0.3   & 13.5 & 1.0  &            &        & $-$6.9   &  1.0 & \tablefootmark{(u)}\\
Camargo 1104   &  5.621   &  $-1.778$   &           &           &            &         & 0.83 & 0.11& 5.4  &  1. 0 &  2.68  &  $-$1.8   & 0.3   & 13.5 & 0.5  &            &        & $-$5.7   &  1.7 & \tablefootmark{(u)}\\
Camargo 1105   &  359.479 &  $2.017 $   &           &           &            &         & 0.90 & 0.11 & 5.8  &  0. 9 &  2.18  &  $-$1.5   & 0.2   & 13.0 & 1.0  &            &        & $-$6.3   &  1.0 &   \tablefootmark{(u)}\\
Camargo 1106   &  357.351 &  $1.683 $   &           &           &            &         &  0.63& 0.11& 4.5  &  0. 4 &  3.54  &  $-$1.5   & 0.3   & 12.5 & 1.0  &            &        & $-$5.7   &  1.6 &   \tablefootmark{(u)}\\    
Camargo 1107  &  357.977  &  0.956 & & & & &0.36 &0.01 &4.0&0.7&4.29&$-$2.2&0.4&13.5&2.0 & & & $-$6.6 & 0.5 &\tablefootmark{(v)}\\
Camargo 1108 & 357.977  & 0.956 & & & & &0.50& 0.01&3.3 & 0.5 & 5.0 & $-$1.80 & 0.3 & 13.5 & 1.5 & & & $-$8.4 & 0.5 &\tablefootmark{(v)}\\
Camargo 1109 & 2.165   &  0.844  & & & & &0.36&0.01 &4.3 & 0.6 & 3.97 & $-$1.50 & 0.2 & 12.0 & 1.5 & & & $-$6.4 & 0.7 &\tablefootmark{(v)}\\
\hline\hline
\end{tabular}
\end{adjustbox}
\tablefoot{
        \tablefoottext{a}{\citet{Garro_2020}}
        \tablefoottext{b}{\citet{Garro2022_Garro02}}
        \tablefoottext{c}{\citet{Garro2021a}}
        \tablefoottext{d}{\citet{2022A&A...658A.120G}}
        \tablefoottext{e}{\citet{Minniti2021_cl160}}
        \tablefoottext{f}{\citet{2022A&A...659A.155G}}    
        \tablefoottext{g}{ \citet{2022A&A...657A..84F}}
        \tablefoottext{h}{ \citet{Gran2021}}
        \tablefoottext{i}{\citet{2011A&A...527A..81M}}
        \tablefoottext{l}{\citet{Fernandez_Trincado_2021}}
        \tablefoottext{m}{\citet{Olivares_Carvajal2022_VVVCL001}}
        \tablefoottext{n}{\citet{2022A&A...657A..67D}}
        \tablefoottext{o}{\citet{Obasi2021}}
        \tablefoottext{p}{\citet{RomeroColmenares2021}}
        \tablefoottext{q}{\citet{Minniti2021_M48}}
        \tablefoottext{r}{\citet{Minniti_2017}}
        \tablefoottext{s}{\citet{2011A&A...535A..33M}}
        \tablefoottext{t}{\citet{Minniti2021a}}
         \tablefoottext{u}{\citet{Camargo_2018}}
         \tablefoottext{v}{\citet{10.1093/mnrasl/slz010}}
         \tablefoottext{z}{\citet{2023arXiv230406904P}}
         \tablefoottext{y}{\citet{2023arXiv231009868G}}
         \tablefootmark{w}{\citet{Minniti_VVVCL002}}
    }
\label{table:Table1}
\end{table*}

\section{Milky Way globular cluster luminosity function}
\label{MWGCLF}
Until now, all the information that we have about the MW GC system were based on the known and well-characterized Galactic GCs, listed in previous compilations such as the catalogs from \cite{1996AJ....112.1487H} and \citealt{Vasiliev2021b}. Discovering and identifying these clusters has been easy over time, as most of them are very bright, massive, highly concentrated, and hosting a larger number of stars. On the other hand, new GCs (Table \ref{table:Table1}) are mainly faint, widespread, embedded in very obscured regions. Therefore, the detection of low-luminosity GCs, possible thanks to powerful surveys, is only able within our Galaxy, while they are not visible in external galaxies. Hence, this may allow us to explore a region of the LF not accessible before, going toward the tail of this feature and increasing statistics at these luminosity ranges.\\

The MW GC LF can be used as a distance indicator \citep{Rejkuba2012} or to constrain the GC history \citep{Kruijssen_2009}.  In our case, we employ it for the second scope.  As mentioned earlier in this paper, many dynamical processes can destroy a GC,  and all these events can have an effect on the MW GC LF, affecting in particular the faint end tail of the distribution. We investigate this in order to see how the known MW GC LF changes if we include all the recently discovered GCs. 

As we can see from Table \ref{table:Table1} and Figure \ref{fig:MW_LF} (orange lines), the LF for the new GCs extends toward much lower luminosities. Actually, we found a double peaked distribution: the first peak at $M_{V}^{new}~=~-6.06\pm0.80$ mag and a second one at $M_{V}^{new}~=~-3.83 \pm 0.62$ mag (this is a tentative one since the $1\sigma$ Poisson error is $\pm 2$ GCs, therefore the statistical significance of this peak is very low). Moreover, we constructed the LF for all known and well-characterized GCs (blue lines), using the $M_V$ values from the 2010 compilation of the \citet{1996AJ....112.1487H} catalog.   A less pronounced double-peak is present for this sample at $M_{V}^{old}= -7.46\pm 0.20$  and $M_{V}^{old}= -4.35 \pm 0.20$, in agreement with the literature (e.g., \citealt{Garro2021b}).  At this point, if we merge the two samples,  thus upgrading the "typical" GC LF (green lines), we find that the distribution is slightly shifted toward lower luminosities, with the bright peak at $M_{V}^{up}= -7.00\pm 1.30$ mag and the faint one at $M_{V}^{up}= -4.10 \pm 0.49$ mag, with an indication of elongated faint end tail. However, the significance and physical meaning of the LF peaks must be taken with caution, because it is well known that the GCLF is asymmetric when plotted in magnitude bins (see, e.g., very similar histograms in \citealt{Fall_2001}). In this instance, the statistical significance of the second peak is not as robust, given the limited number of clusters within that range (specifically, at $-5<M_V<-3$, there are 26 objects with Poisson error for each bin of $\pm 2-3$); however it does not disappear even when we include the Poisson error. Hence, this observation might imply the existence of numerous other faint clusters yet to be discovered, likely obscured by the considerable challenges in their detection. This study could serve as an initial step in delving into the analysis of this faint segment of the MW GC LF.\\
Additionally,  we can note that the uncertainties on the GCLF peak magnitudes seem large, even for the brighter peak, when instead other studies usually find uncertainties of 0.1-0.2 mag (as reviewed, for example, by \citealt{Rejkuba2012}). However, in our case, the uncertainties are greater due to the large uncertainties in the luminosities of each cluster (refer to Table \ref{table:Table1}). This is attributed to the methodology used in calculating luminosities in prior literature. Colour-magnitude diagrams indicate that main sequence stars are missing in their compilation, resulting in lower-limit luminosity values. Consequently, this leads to significant uncertainties, thereby contributing to higher uncertainties in the MW LF peak. 

As demonstrated by several works, the GC LFs is not Gaussian nor universal for all galaxies \citep{Rejkuba2012}. The shape of the GC LF deviates from the Gaussian symmetric distribution, since it shows a longer tail toward the faint end. This trend is related to the main dynamical processes that can destroy a GC (which can be amplified especially when galaxy merger events are involved), which are: {\it (i)} the mass loss due to stellar evolution, such as the supernova explosions \citep{Lamers2010}; {\it (ii)} dynamical friction \citep{Alessandrini2014}; {\it (iii)} tidal shock heating by passages through the bulge or disk \citep{Piatti_Carballo2020}; and {\it (iv)} evaporation due to two-body relaxation \citep{Madrid2017}. Moreover, \cite{Kruijssen_2009} suggested that the efficiency of the dynamical evolution are linked to the initial cluster LF and mass function, which depend on the M/L parameter differently.  In turn, M/L depends on the cluster mass dependent mass-loss rate (dissolution time $t_{dis}\propto M^{0.7}$ and $t_{dis}\propto t_{rh}^{0.75}$, where $t_{rh}$ is half-mass relaxation time, \citealt{Kruijssen_2009}).  This indicates that if M/L increases with mass and luminosity, the low-mass, and low-luminosity GCs, present in our compilation, are more strongly depleted in low-mass stars. This because the massive stars are more segregated in the central regions, whereas the low-mass stars are preferentially located in the cluster's outskirts, where the cluster's potential well is less strong than the central regions; thus, the chance of losing stars is most probable in the external regions of the cluster.  Hence, the estimation of the luminosity, which we are using (see Table \ref{table:Table1}, and reference there in) is reliable, since it mainly comes from evolved and brighter stars. Nonetheless, a dependence of M/L on mass is not obvious, since it is still not the impact of the mass function can change depending on the cluster's metallicity, as suggested by \cite{10.1093/mnras/stw2488}.\\ 

Another important feature that we have to include is the fact that the MW GC LF may be bimodal, even if the second peak does not contain a high number of clusters. However, we want to compare the obtained MW GC LF with the M31 galaxy's LF (\citealt{Peacock2010, 10.1093/mnras/stu771, Garro2021b}) -- which shows a double peaked distribution --  to see if we can deduce some important clues about our Galaxy. By this comparison, we can deduce that the second peak may be related to GCs, which have been accreted or survived disruption processes \citep{10.1093/mnras/stu771}. This may be supported by the fact that the new clusters that are very faint ($M_V>-5.0$), they are also younger than 11 Gyr, more metal-rich than -1.0, and located in the bulge, where dynamical processes are very strong (Figure \ref{fig:newimage}). Precisely, Ferrero 54 and Gaia 2 located at larger $R_G$ are disk-like GCs, and Patchick 125 is the ancient clusters in our compilation. If this is true, our targets have lost a high percentage of their stars, therefore likely those low-luminosity (and low-mass) GCs may be what remains of more massive progenitors. This result is in agreement with recent studies that suggest that the Galactic stellar halo has assembled from tidally disrupted dwarf galaxies (see e.g., \citealt{10.1093/mnras/sty982, Conroy2019,Kruijssen2019}). However, most of the stellar streams, or the accretion events in general, have been discovered across the Galactic halo \citep{Ibata_2001, 2001ApJ...548L.165O,10.1093/mnras/stad321}. This may represent a "selection effect," given that studying the halo is easier than exploring the bulge, just because the stellar crowding and extinction are irrelevant there. Finding new objects in the bulge, which may be what remains of disrupted GCs, and considering why they were not accreted, could pave the way for the study of ex situ populations also in the bulge. Nonetheless, we need numerical simulations and spectroscopic analysis that can confute this hypothesis. For that, we can speculate that this new MWGC LF may be the result of different processes, such as a higher number of accreted objects, efficient destruction due to the above-mentioned dynamical processes that occurred both in the Galactic halo and in the bulge, and/or incompleteness of our catalog since many other faint and obscured objects still remain to be identified. In particular, we suspect that with the discovery and confirmation as real GCs of other candidates, which are also of low luminosity, it may be possible that the LF can change once completeness is achieved. 

{
\sidecaptionvpos{figure}{c}
\begin{SCfigure*}
\centering
\includegraphics[width=6cm, height=6cm]{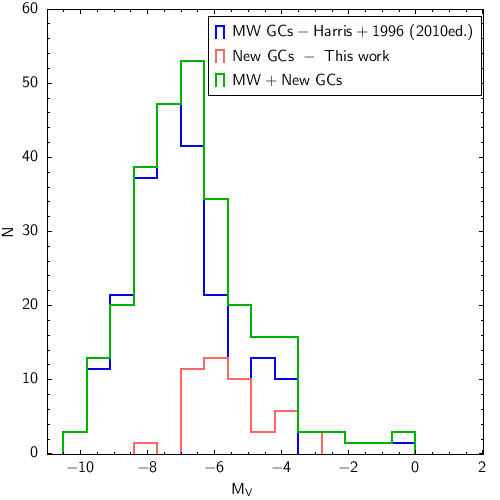} 
\includegraphics[width=6cm, height=6cm]{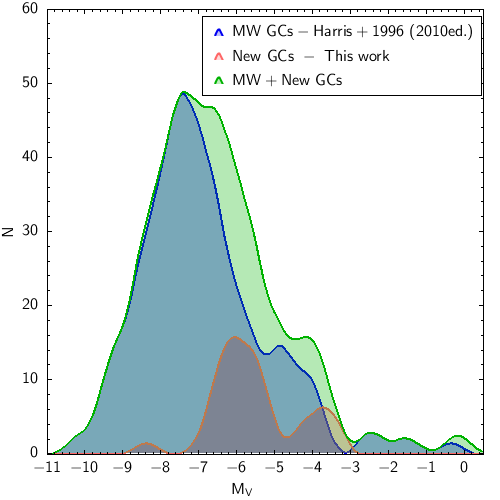}
\caption{Luminosity functions (LFs).  The left panel shows three histograms depicting the LFs for the known MW GC system by \citet{1996AJ....112.1487H} (2010 edition), using the dashed blue line, for the GCs analyzed in this work and listed in Table \ref{table:Table1}, using a dashed orange line and the sum of the two mentioned distributions in the green line. In the right panel, we show the LFs using the KDE technique for the three samples.}
\label{fig:MW_LF}
\end{SCfigure*}
}

\begin{figure*}
\centering
\includegraphics[width=18.5cm, height=6cm]{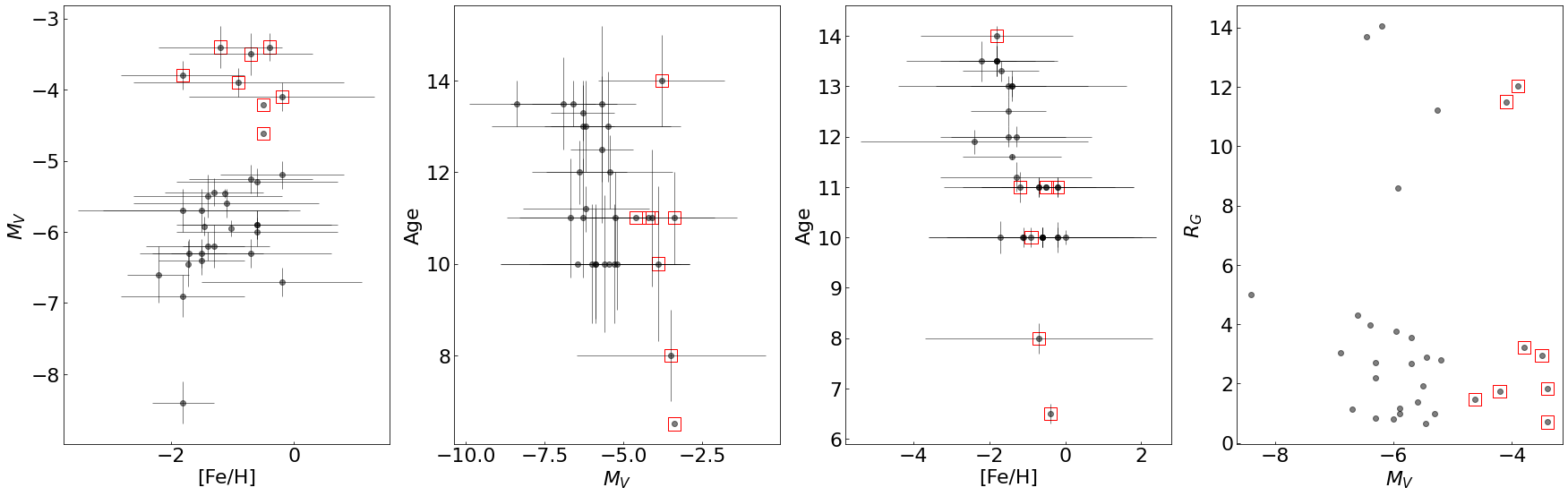} 
\caption{[Fe/H]-$M_V$ (left panel), $M_V$-Age (middle-left panel), [Fe/H]-Age (middle-right panel), and $M_V$-$R_G$ diagrams for the new GC sample. With red squares, we selected only very faint clusters with $M_V>-5.0$. }
\label{fig:newimage}
\end{figure*}

\section{Metallicity distribution}
\label{MDF}
\begin{figure*}[htpb]
\centering
\includegraphics[width=6cm, height=6cm]{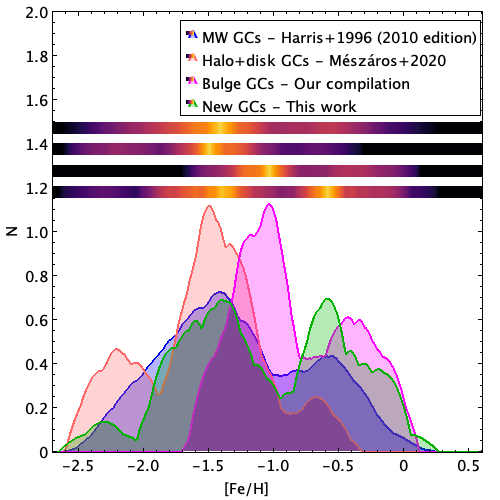} 
\includegraphics[width=12cm, height=6cm]{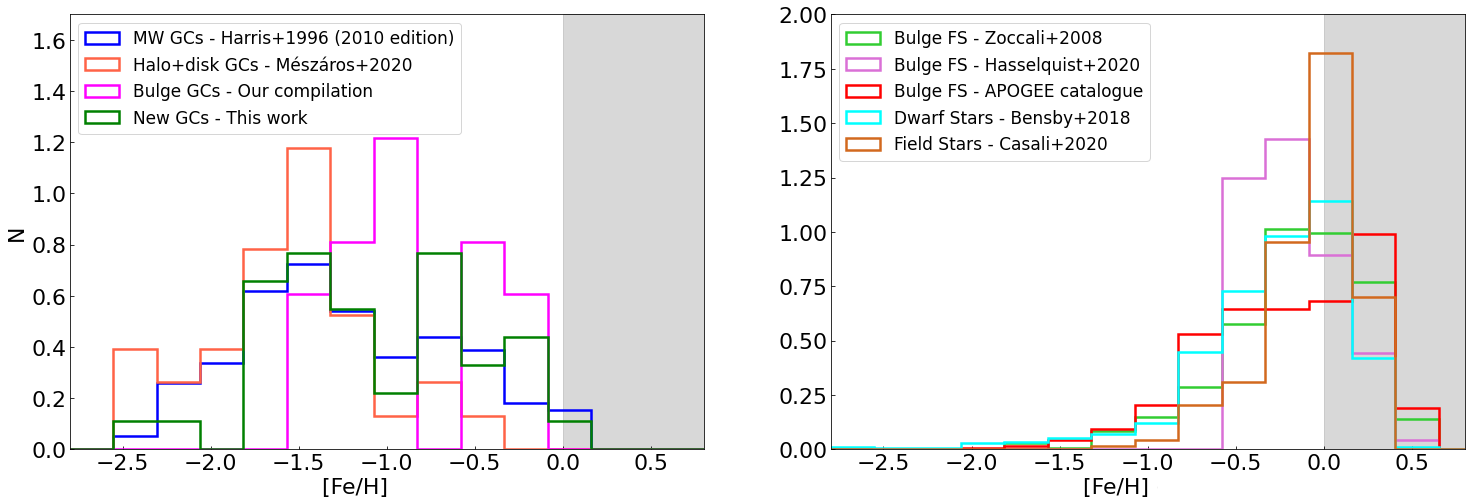} 
\caption{Metallicity distributions.  Right and in the middle panels we show the KDEs and histograms for the known MW GCs for the \citet{Meszaros_2020} (orange lines), \citet{1996AJ....112.1487H} (2010 edition - blue lines),  bulge GCs (Table \ref{table:bulgeGCs} - magenta lines), and our work (green lines) samples, respectively. In the right panel, we also show the densograms in order to better discriminate the higher peaks (yellower colors) of the distributions. On the left panel, we show histograms for different samples of bulge FSs, as specified by the legend. The grey areas, in the middle and left panels, are used to point out that at metallicity [Fe/H] $>0$ we do not have any GCs, while a higher percentage of bulge FSs is located at that range.  }
\label{fig:MW_MD}
\end{figure*}

The metallicity distribution (MD) is an important tool for obtaining information about the age and formation history of a stellar system, especially when this is investigated in different environments.  
As shown in Figure \ref{fig:MW_MD}, we compared the new GC MD (green line) with the MDs of known GCs using three samples: the 2010 compilation of the \citet{1996AJ....112.1487H} catalog (blue line),  the \citet{Meszaros_2020} catalog (orange line), which includes, in particular, the halo+disk GCs, and our compilation of bulge GCs listed in Table \ref{table:bulgeGCs}. The first sample contains 157 GCs and their [Fe/H] values are both spectroscopic and photometric estimates\footnote{Extensive references are given for the Harris' catalog at \url{https://physics.mcmaster.ca/\%7Eharris/mwgc.ref}}, whereas the second includes 31\footnote{\citet{Meszaros_2020}  excluded from their catalog GCs exhibiting significant differential reddening (E(B-V) $>0.4$), which introduces errors in estimations of atmospheric parameters and, consequently, the chemical abundance estimates.} known GCs and the [Fe/H] is taken from SDSS-IV APOGEE-2, while the third catalog is created using [Fe/H] and [Mg/Fe] taken from the literature, as specified in the fourth column. \\
\citet{Meszaros_2020} found that the [Fe/H] metallicities that they derived are on average 0.162 dex higher than those from the Harris catalog, so we subtracted this quantity from their [Fe/H]  to bring the MDs to the same scale. 
Due to the fact that the histograms depend on the bin size, we also preferred to construct a kernel density estimation (KDE) diagram, using the {\ttfamily epanechnikov} kernel, a smoothing of 0.25, and normalized per area. The KDE code provides us the highest peak of the metallicity distributions.
Independently,  to better identify the peaks of the distributions, we overplotted a densogram for each sample, using the {\ttfamily acos} scaling\footnote{We constructed the densogram with {\ttfamily TOPCAT}, and we used the {\ttfamily acos} scaling only because the peaks of the distributions were more marked than the {\ttfamily linear} scaling.}   in agreement with the previous method.  A third and more robust method that we implemented is the Gaussian mixture model (GMM). We tested different Gaussian components and that the best fit came from  a two-Gaussian fit, from which we obtained the following peaks and errors. In addition, these results are in agreement within the errors with the previous estimations. In all cases, we found double-peaked distributions (see Figure \ref{fig:MW_MD}):
{\it (i)} for the \citet{Meszaros_2020} sample, the metal-poor peak is at [Fe/H] $=-1.49\pm 0.40$ dex and the metal-rich peak is at [Fe/H]~$~=-0.56\pm 0.46$ dex.  The latter is not clearly visible in the histogram distribution,  due to the choice of bin size, but it is clear in the KDE diagram. Finally, a third peak at very metal-poor range is detectable at [Fe/H]~$~=-2.20\pm 0.38$ dex;
{\it (ii)} for the \citet{1996AJ....112.1487H} (2010 edition) sample, the metal-poor peak is at [Fe/H] $=-1.59\pm0.36$ dex and the metal-rich peak is at [Fe/H]~$~=-0.59\pm 0.35$ dex, in agreement with the literature (e.g., \citealt{Minniti1995,Carretta2009, Smith_2000,ness_freeman_2016, Dias2016,Garro2021b});
{\it (iii)} for the bulge GCs sample (see Table \ref{table:bulgeGCs}),  the metal-poor peak is at [Fe/H] $=-1.03\pm 0.20$ and the metal-rich peak is at [Fe/H]~$=-0.38\pm 0.20$;
{\it (iv)} for the new sample containing clusters in Table \ref{table:Table1}, metal-poor peak is at [Fe/H] $=-1.38\pm0.35$ dex and metal-rich is at [Fe/H]~$~=-0.59\pm 0.23$ dex, with another tentative peak at [Fe/H]~$~=-0.12\pm 0.20$ dex.

\begin{table}[htpb]
\centering 
\footnotesize
\renewcommand{\arraystretch}{1.1}
\caption{Parameters for bulge GCs. We construct our compilation of 20 bulge GCs, collecting [Fe/H] values from the literature.}
\begin{tabular}{lcl}
\hline\hline
Cluster ID        &  [Fe/H]&      References\\
\hline
Terzan  2    &  $-0.85\pm 0.04$   & \cite{Geisler2021_CAPOS}    \\
Terzan 4    &  $-1.40\pm 0.05$   & \cite{Geisler2021_CAPOS} \\
HP 1        &  $-1.20\pm 0.10$   &                 \cite{Geisler2021_CAPOS}     \\
Terzan 9    &  $-1.40\pm 0.16$   &                \cite{Geisler2021_CAPOS}  \\
Djorg 2     &  $-1.07\pm 0.09$   &               \cite{Geisler2021_CAPOS} \\
NGC 6540    &  $-1.06\pm 0.06$   &                \cite{Geisler2021_CAPOS} \\
NGC 6642    &  $-1.11\pm 0.04$   & 
\cite{Geisler2021_CAPOS}                \\
NGC 6558    &  $-1.17\pm 0.01$   &\citet{Barbuy2018}             \\
NGC 6522    &  $-1.05\pm 0.20$   &   \cite{Barbuy2021_ngc6522}                    \\
NGC 6316    &  $-0.50\pm 0.06$   &        \cite{Dias2016}               \\
NGC 6355    &  $-1.46\pm 0.06$   &         \cite{Dias2016}                \\
NGC 6401    &  $-1.12\pm 0.07$   &       \cite{Dias2016}                     \\
Palomar 6   &  $-0.85\pm 0.11$   &    \cite{Dias2016}                  \\
NGC 6440    &  $-0.24\pm 0.05$   &      \cite{Dias2016}            \\
NGC 6528    &  $-0.13\pm 0.07$   &      \cite{Dias2016}                 \\
NGC 6539    &  $-0.55\pm 0.06$   &      \cite{Dias2016}                     \\
NGC 6553    &  $-0.13\pm 0.01$   &   \cite{Dias2016}            \\
NGC 6441    &  $-0.46        $   &\cite{Carretta2009}, \cite{1996AJ....112.1487H}    \\
NGC 6171    &  $-0.87\pm0.13 $   &   \cite{Meszaros_2020}                          \\
NGC 6388    &  $-0.41\pm0.16 $   &  \cite{Meszaros_2020}                           \\
\hline\hline
\end{tabular}
\label{table:bulgeGCs}
\end{table}

Basically, the distributions, obtained using these different samples, are in agreement between them, since both metal-rich and metal-poor peaks show similar values. 
In this way, we can confirm the metallicity bimodality of GCs as expected, since it is well studied in the literature (e.g., \citealt{Tonini2013}). However, we can use the MDs in order to {\it (i)} understand something more about our compilation and {\it (ii)} apply known models in order to draw conclusions regarding the nature of our sample. \\

\begin{figure}[htpb]
\centering
\includegraphics[width=6cm, height=5cm]{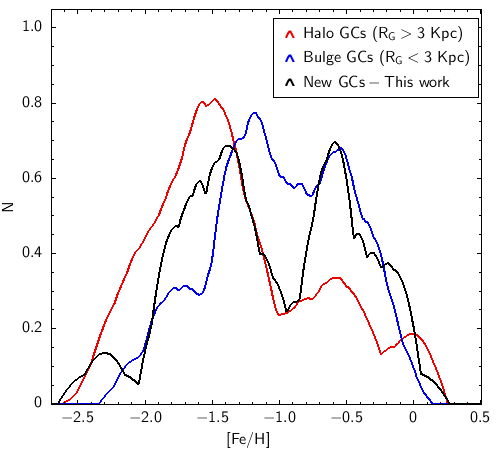} 
\caption{Same details as in Figure \ref{fig:MW_MD} (right panel). We split the \citet{1996AJ....112.1487H} (2010 edition) GC
catalog into inner (bulge) GCs located at $R_G<3$ kpc (red line) and outer (halo) GCs at $R_G>3$ kpc (blue line). The new GCs sample is drawn in black line. We use this comparison to better understand the separation between bulge and halo clusters and to show that new GCs sample follow
a similar trend as the inner GCs.}
\label{fig:new}
\end{figure}

Figure \ref{fig:new} displays that the new GCs sample, which are mainly situated in the MW bulge, follows the same trend as the inner (bulge) GCs. In addition, we can see a farther tentative peak at [Fe/H] $\approx -0.1$. This may suggest that our sample may be contaminated by younger stellar systems, which should be more metal-rich than older population, as GCs are. This might be expected given that age estimates are approximate for all our targets, and OCs may be included in our sample. Examples could be: Ferrero 54, which was classified as MW disk GC by \cite{Garro2023}, with an age $=11$ Gyr and [Fe/H] $=-0.2$; and Kronberger 49 located at $R_G = 1.14$ kpc, with an age $=11$ Gyr and [Fe/H] $=-0.2$. For that reason, we need both spectroscopic analysis and deeper CMDs to constrain their parameters.\\  

Moreover, comparing all distributions, we can see that the number density of the metal-poor part is in better agreement, whereas the metal-rich one is shifted to higher values for the new and bulge GC samples than the other two. We can expect this effect because the metal-rich component is concentrated more toward the inner regions of the Galaxy, where the majority of our new targets are located, while the metal-poor one becomes more significant at larger Galactocentric distances. Indeed, as found by \cite{Tonini2013} model (their Figure 2), the metal-rich distribution was formed locally in the main progenitor at $z \sim 2$, while the metal-poor component was formed in satellites at an epoch $z\sim 3-4$, and then merged with the main progenitor. Therefore, connecting to the discussion made in Section \ref{MWGCLF}, and comparing our distribution with the Tonini's model, the metallicity bimodaly is a direct prediction of a hierarchical clustering scenario. In consequence, we argue that the metal-poor component is mainly the contribution due to the tidal disruption of dwarf-like objects, whereas the metal-rich population is related to the formation of the bulge and/or disk (see e.g., \citealt{Mouhcine2006}). This may imply that many of our targets may be formed in situ, and they might be classified as main progenitors, following the nomenclature from \cite{2019A&A...630L...4M}. \\
Moreover, the height of the MD's peaks can give us clues about the star formation histories. Figures \ref{fig:MW_MD} and \ref{fig:new} show that metal-poor and metal-rich peaks are similar, especially if we include in the total compilation the new GC sample. Therefore, based on Tonini's model, this trend suggests that the MW experienced a balanced star formation ($M_{SF}/M_0 = 0.2$) and merger history ($M_1/M_0 = 0.3$)\footnote{Where $M_{0}$ is the final stellar mass, $M_{SF}/M_0$ star formation history parameter, and $M_1/M_0$ is assembly histories parameter.}. Moreover, we know that multiple components, with different ages and metallicity, co-exist in the bulge, suggesting that star formation does not stop at $z\sim 2$; father, the final stellar mass is formed at later (or $z<2$) time, hinting that younger (e.g., age of $\sim 10$ Gyr) GCs may have formed later in the MW bulge, due to an increasing in gas and so a burst in star formation due to merging events. Once the age of our targets has been confirmed, our young candidates might belong to this younger populations, and this can help confirm whether the formation of GCs may occur later in the MW bulge, indicating a strong presence of gas at that stage; otherwise this could be linked to accretion events, which enriched our Galaxy and formed a new generation of star clusters. However, this is a speculation until this statement is confirmed observationally as well as via simulations and models.   \\

Finally, we want to highlight the fact that spectroscopic analysis should be performed to recover orbital parameters and chemical abundances, thus constraining the real nature of these systems. Furthermore, it is likely that many other GCs are missing from our compilation.

\section{Comparisons with field stars}
\label{sec:comparison}
Another important comparison we carried out within the framework of this study considers GCs and FS MDs (see Figure \ref{fig:MW_MD}). In order to do this, we took into consideration five bulge FS public samples by \citet{Zoccali2008, Bensby2017, Hasselquist2020, Abdurrou2022}. For the \cite{Abdurrou2022} catalog, we applied the typical APOGEE selection cuts: {\ttfamily apo\_ps41} $<3.5$, {\ttfamily ASPCAPFLAG} $=0$, {\ttfamily STARFLAG} $=0$, {\ttfamily SNR} $>70$, {\ttfamily LOGG} $<3.6$, {\ttfamily TEFF} $<5500$, {\ttfamily VSCATTER} $<1$, and {\ttfamily ruwe}~$<1.4$. Also, for comparison, we included  Galactic disk FSs, using the catalog from \cite{Casali2020}.   \\
Even though they come from widely different surveys (this was done intentionally since we need high statistics), these are all in good agreement, exhibiting metallicity peaks between -0.5 and 0.0, with a long tail toward the metal-poor regime. In addition, we can see that both bulge and disk FS catalogs seems to follow the same behavior. From Figure~\ref{fig:MW_MD}, we can see that $\sim 30- 40\%$ of stars in all those catalogs show metallicities higher than Solar value, while a very low percentage of stars have [Fe/H] $<-1.0$. Even though a large number of very metal-rich stars have been observed, we still do not know of any GCs in the super-Solar metallicity regime today.  So, we consider the question of where the super metal-rich bulge GCs reside and whether they can feasibly form and exist.  Alternatively, we explore whether there might be a physical process that preferentially destroyed them in the past. \\

To gain clarity in these areas, we need to understand if there is a connection between GCs and isolated FSs. We could resort to a range of possible explanations, such as different star formation scenarios that may have occurred for these two populations, in addition to scenarios related to the formation of the bulge and disk during the early phases of the Universe \citep{Costantin2021}. Evolutionary effects, such as cluster disruption, mass loss, interactions, merging events may also play a role \citep{2017RNAAS...1...16M,Minniti_2019}.  
It is likely that all of these properties and effects have had an effect on the ratio of clusters to FSs at a given metallicity range and location within the Galaxy, and may be responsible for these observed differences.  To date, many studies were focused on the Galactic halo, finding a good agreement between observations and simulations. For instance,  \citet{2020MNRAS.493.3422R}, using an E-MOSAIC simulation,  found that GCs play a sub-dominant role in the build-up of the stellar halo.  Additionally, these authors found that the disruption of GCs contributes between $0.3-14\%$ of the bulge mass, in agreement with observational estimates by \cite{Hughes2019a}. It was also predicted that stars from disrupted GCs in the Galactic bulge show a higher fraction around [Fe/H] $\approx -1$.  A similar fraction was found by  \citet{Schiavon2017} when analysing FSs in the MW bulge in a specific metallicity range of [Fe/H] $<-1$ and, by finding nitrogen-enriched stars, concluded that $14\%$ of the stellar mass of the bulge came from disrupted GCs.
Therefore,  the contribution of disrupted GCs to the stellar population of the bulge constitutes only a small percentage. Thus, we suspect that this factor alone cannot explain this inconsistency in metallicity between clusters and bulge FSs; it is more likely that the ratio of FSs to GCs varies strongly in different environments. For instance, as found by \cite{Lamers2017} the difference between the GC and FS metallicity distributions is not specific to our Galaxy, but is a well known phenomenon in external galaxies as well and it changes if outer or inner regions are taken into account. Reporting what they found: (i) in dwarf galaxies the population is generally metal-poor, but the GCs are usually more metal-poor than FSs; (ii) in the M31 inner and outer halo, the cluster-to-star ratio is high at low metallicity and low at high metallicity; (iii) in both the inner and outer regions the relative contribution of the clusters at low metallicity is larger than that of the stars; (iv) in both the inner and the outer regions, the relative contributions of the clusters at high metallicity is smaller than that of the stars; and (v) finally, the ratios of Fornax dwarf spheroidal galaxy are more similar to those in the outer regions of the larger galaxies than to the inner regions.

\section{Age-metallicity distribution}
\begin{figure}[htpb]
\centering
\includegraphics[width=9cm, height=9cm]{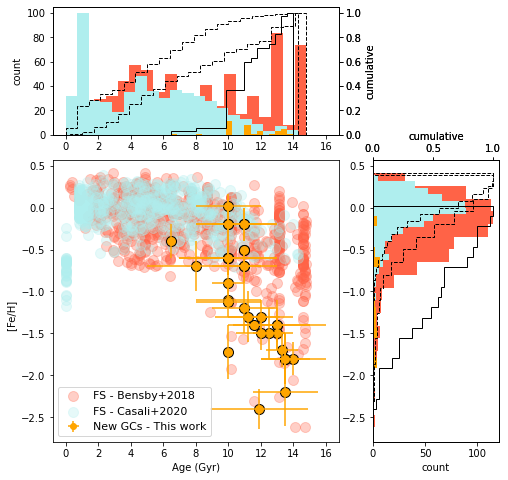} 
\caption{Age-metallicity distribution for FSs (using catalogs from \citet{Bensby2017} as red points  and \citet{Casali2020} as cyan points) and for new GCs (orange points).  In the top and right panels, we show the respective histograms and cumulative distributions for FSs (red and cyan histograms and dashed black lines,  respectively) and for new GCs (orange histograms and continuous black lines). }
\label{fig:Age_Met}
\end{figure}

Another distribution that we want to take into consideration in order to understand the main differences between GCs and FSs is the age-metallicity distribution (Figure  \ref{fig:Age_Met}).  We used the catalogs  by \citet{Bensby2017} (red points) and \citet{Casali2020} (cyan points), while we considered only the new GCs (orange points) by the present compilation, since (in general), these are low-luminosity and low-mass clusters, so we foresee that they have lost a fraction of stars during their life (see Section \ref{MWGCLF}). As expected, we can see that there is an anti-correlation between age and metallicity, since the age decreases with metallicity. In particular, we may note that the younger and metal-rich population is represented by FSs, while GCs occupy the right-bottom part of the diagram, with lower metallicities and older ages.  This is also highlighted by the histograms and cumulative distributions displayed by Figure \ref{fig:Age_Met}.  GCs are more segregated at age $> 10$ Gyr and their [Fe/H] spans over all values, especially at [Fe/H] $< -0.5$. In our compilation, the only two clusters with age $<10$ are VVVCL002 (age $>6.5$ Gyr) and Patchick 126 (age $>8$ Gyr). For the first, \cite{Minniti_VVVCL002} demonstrated that VVVCL0002 is an old GC, since its high $\alpha$-enhancement ($[\alpha/Fe]>+0.4$) is connected to rapid chemical evolution dominated by core-collapse supernovae rather than type Ia supernovae, formed together with other clusters and field stars present today in the Galactic bulge, rather than a younger open cluster or the remnants from an (already-disrupted) dwarf galaxy. Patchick~126 is already classified as an in situ GC by \cite{2023MNRAS.tmp.3751B}, suggesting that its formation occurred  $\approx 11.7 - 12.7$ Gyr ago, according to in situ GC ages. Therefore, we keep these lower limits, even if their ages seem to be older than those listed in Table \ref{table:Table1}. On the other hand, FSs are more segregate at [Fe/H] $> -0.7$, while their ages span over a wide range of values including intermediate age with age $< 6$ Gyr. 

\section{Discussion}
\label{discussion}
The literature extensively addresses differences in metallicity distributions between FS and GC populations, not only within our Galaxy, but also in external galaxies. For example, \cite{1981ApJ...245L...9F} proposed that the color variations observed in the spheroidal components of external galaxies (NGC~4374, NGC~4406, NGC~4621, and M87) compared to their GC systems suggest a disparity in metallicity, indicating that GCs are more metal-poor than the spheroidal components. This implies a distinct chemical enrichment history for the GC system and the spheroidal component, suggesting that GCs are likely form a component that is dynamically and chemically distinct from the spheroid.
More recently, \cite{2012A&A...544L..14L}, in their study of the Fornax dwarf spheroidal galaxy's GC system and its metal-poor FSs, suggested that the primary distinction in the metallicity distribution between FSs and GCs arises from the ability of each Fornax cluster to accrete a substantial amount of gas with the same composition as the first-generation stars. In contrast, limited gas of this composition was available for the formation of additional FSs. In this section, we want to sum up and attempt to understand whether there is indeed a verified link between FSs and GCs. We also  explore how this might help answer the questions posed in Section \ref{sec:comparison}. \\

One caveat is related to the fact that our GC compilation may still be incomplete, as many other star clusters could remain hidden, due to crowding and extinction. Furthermore, we predict that the chemical analysis of these clusters will be  performed in order to confirm or refute the nature of the new GCs included in the present work, as well as to better distinguish between the natures of GCs and OCs. Finally, we need further support from numerical simulations to understand whether it is possible to form super-metal-rich GCs, understand their probable evolution until their extinction, or (on the other hand) to understand why it is not possible to form them. \\

However, we find it surprising that there are a high percentage of super metal-rich FSs, but no  GCs in the same metallicity range. For instance, presuming that we are roughly counting that there are $30-40\%$ of stars with [Fe/H] > 0, out of $\sim 50$ bulge GCs, we would expect $\sim 20-25$ to have super-Solar metallicity; however, there are none. This fact is independent of how the bulge GC is defined, namely: because there is no definitive classification, the criteria used to describe them can differ among works in the literature.  Again, if we use another sample of fewer bulge GCs, with only 30 bulge GCs in total,  there is still not a single super-Solar metallicity GC present. Therefore, we foresee that the answers to all our questions must be sought in the evolutionary difference between these two stellar populations. This is supported by the fact that we found significant differences, as displayed in Figure \ref{fig:Age_Met}, where we recognise three evolutionary moments.  We can suppose that young (age $<5-6$ Gyr) and (super) metal-rich FSs may be the result of a dissolution of old open clusters (OCs), since they are younger, more metal-rich, less massive, less bound, with shallower potential wells and therefore with shorter break-up times than GCs (with the caveat that the OC-GC boundary is not so well defined).  This may be supported by the fact that many OCs can disintegrate, as recently suggested by \citet{2022MNRAS.517.3525B}. In fact, they found that many OCs show extended features, such as tidal tails and extended coronas. \\

Similar features have also been discovered in MW GCs. For instance, \citet{2019MNRAS.483.1737K,2021A&A...645A.116K} concluded that some GCs analyzed in the MW halo show the presence of extra-tidal RR Lyrae stars. So, we can expect (and attempt to explore in the future) the notion that several small and low-luminosity GCs  analyzed here may be exhibit similar extra-tidal features, as found by \citet{2023arXiv230105166F} in other Galactic GCs (e.g., NGC 3201, NGC 4590, NGC 5466, Pal 5). However, if a GC's extra-tidal features are composed of stars such as RR Lyrae, these, in particular, contribute with old and metal-poor stars. Therefore, if we speculate on the fact that (super) metal-rich FSs are the result of OC dissolution or groups of stars formed later from [Fe/H]-enriched gas,  we probably do not see super metal-rich GCs because at the time when GCs formed (10-13 Gyr ago), the gas did not have enough time to get rich in metals to form super metal-rich GCs. Thus, we suggest that they never formed in the first place. \\

On the other hand, \citet{2023A&A...674A.148N} also modeled GC destruction for objects with orbits that bring them close to the Galactic centre, finding that the least massive clusters are quickly ($<10$ Myr) destroyed. Instead, the most massive clusters have a lengthy evolution, displaying variations in the morphology, especially after each passage close to the supermassive black hole.
Therefore, in order to explain why we have not yet detected super metal-rich GCs, we cannot exclude other options: {\it (i)} we have not detected them yet and that may be because they live in very crowded and reddened regions where the separation from FSs is impossible or {\it(ii)} they have been destroyed, because they lived in preferentially dangerous regions, or they have orbits that are prone to efficient dynamical effects; alternatively, {\it(iii)} they did not form, because of some unknown reason, {\it(iv)} another possibility is the accretion of metal-rich elliptical galaxy, such as M32, which is not known to contain GCs \citep{2009AJ....138.1985R}; or {\it(v)} finally, a combination of these tentative answers may serve as an apt approach to explaining why super metal-rich GCs are not observed today. \\

\section{Conclusion}
Several dozens of new Galactic GCs have been discovered in the past decade or so thanks to the advent of new large surveys in the optical and near-IR, significantly augmenting the MW GC system. Therefore, our aim is to collect them in a single catalog to show that, in reality, our Galaxy could potentially host more than 200 GCs. Taking into account all the limitations described in Section \ref{targetselection}, this work allowed us to update the data on the MW GC system, adding 37 new GC candidates; in addition,  we were able to upgrade the LF and MD data. Independently, \cite{2024arXiv240503068B} analyzed a similar updated sample of GCs in the bulge, arriving at analogous conclusions. 

One of the main achievements of our work is the fact that we are exploring the faint tail of the MW GC LF. The "updated" GC LF is shifted toward slightly lower luminosities and may indicate a double-peaked distribution. When comparing the MW GC LF with that of Andromeda's GC LF and assuming the authenticity of the second peak, it suggests that the MW may harbour accreted objects. Alternatively, these objects could be survivors of a dynamical process that led to their dissolution or disruption. Due to the fact that their age estimates are approximate, we can only speculate on the fact that some of these low-luminosity objects could be what remains of those objects that today the {\it James Webb} Space Telescope is observing at high redshift (see Section \ref{Introduction}). Alternatively, there may be a connection between those compact red objects and diffuse Galactic GCs, present in our compilation. Moreover, the MD shows a bimodal distribution, with a third tentative peak. We find that new GCs follow the same behavior as the bulge GC population (as shown in Figure \ref{fig:new}), as expected due to their location inside the MW. Furthermore, it seems to be accepted that bulge GCs may be the first stellar objects formed during the formation of the bulge and/or disk (an example may be Liller 1 - \citealt{Dalessandro_2022}), while the metal-poor GCs, located at higher Galactocentric distances, may represent the accreted component (e.g., \citealt{Forbes_Bridges2010} and reference therein). Nevertheless, we cannot exclude the possibility that the Galactic bulge (as well as the Galactic halo) may contain ex situ clusters or those formed in a second moment from accreted gas, supporting the scenario that the bulge was formed first by monolithic collapse and then grew up hierarchically via merging galaxies. This is a hypothesis that must be confirmed spectroscopically. Furthermore, no GCs with super-solar metallicity have been discovered so far. This is perhaps the most important conclusion. To obtain clues to resolving the questions we consider in Section \ref{sec:comparison}, we definitely need the support of: {\it (i)}  spectroscopic analyses to confirm the nature of these GCs, derive their chemistry, and reconstruct their orbits; {\it (ii)} numerical simulations that can help quantify how many super metal-rich GCs we can expect in the MW, if they can or cannot exist, if  they were destroyed, and if so, to understand why and how; and, finally, {\it (iii)} deeper observations in order to reach the MSTO and derive robust ages. \\

The data and conclusions presented in this work are compiled on the basis of existing surveys, such as VVV-VVVX, 2MASS, and SDSS in the near-infrared (NIR), as well as the {\it Gaia} mission in the optical wavelengths. Hence, more progress is expected to come with the advent of future instrumentations such as LSST, MOONS, 4MOST, and others.

\begin{acknowledgements}
The authors are grateful to the anonymous referee for providing helpful comments and suggestions, which have improved the content of this paper.
We thank Dr. Anna Barbara de Andrade Queiroz and Dr. Bruno Dias for their invaluable suggestions. \\ We gratefully acknowledge the use of data from the ESO Public Survey program IDs 179.B-2002 and 198.B-2004 taken with the VISTA telescope and data products from the Cambridge Astronomical Survey Unit.  This work presents results from the European Space Agency (ESA) space mission {\it Gaia}. {\it Gaia} data are being processed by the {\it Gaia} Data Processing and Analysis Consortium (DPAC). Funding for the DPAC is provided by national institutions, in particular the institutions participating in the {\it Gaia} MultiLateral Agreement (MLA). The {\it Gaia} mission website is \url{https://www.cosmos.esa.int/gaia}. The {\it Gaia} archive website is \url{https://archives.esac.esa.int/gaia}. \\

E.R.G. gratefully acknowledges support from ANID PhD scholarship No. 21210330 and ESO Fellowship program.  D.M. gratefully acknowledges support from the ANID BASAL projects ACE210002 and FB210003, from Fondecyt Project No. 1220724, and from CNPq Project 350104/2022-0. J.G.F-T gratefully acknowledges the grant support provided by Proyecto Fondecyt Iniciaci\'on No. 11220340, and from the Joint Committee ESO-Government of Chile 2021 (ORP 023/2021). 
\end{acknowledgements}

\bibliographystyle{aa.bst}
\bibliography{references_thesis}

\end{document}